\def\et{$E_T$}                          %ET
\def\met{\mbox{${\hbox{$E$\kern-0.6em\lower-.1ex\hbox{/}}}_T$}} %missing ET
\def\D0{D\O}                            %D0
\def\ppbar{$p\overline{p} $}            %ppbar
\def\etal{{\sl et al.\ }}
\def\etaj2lt{$|\eta_2| <$}
\def\gtetaj2lt{$< |\eta_2| <$}

\def\wjets{\mbox{$W +$ Jets}}
\input psfig
\documentstyle[preprint,psfig,aps,floats,tighten,subfigure]{revtex}
\begin{document}
\title{A SUMMARY OF RECENT COLOR COHERENCE RESULTS}
\author{
Nikos~Varelas                             \\
{\em Department of Physics, University of Illinois at Chicago, Chicago,
Illinois 60607, USA}
}
\maketitle
\baselineskip=14.5pt
\begin{abstract}
Recent experimental results on color coherence phenomena from
$e^+e^-$, $ep$, and \ppbar\ collisions are presented.  The data are compared
to analytic perturbative QCD calculations based on the modified 
leading logarithm approximation and the local parton hadron duality
hypothesis.  
\end{abstract}
\baselineskip=17pt
\section{Introduction}
\indent

An important goal in the study of high energy hard collision properties is the
detailed understanding of the hadronic final state and its characteristic jet
structure.  Perturbative QCD (pQCD) calculations have been used to describe the
production of jet final states.  However, the jet structure still relies on
phenomenological models to describe how the partonic cascade evolves into the
final state of hadrons.

In the picture implemented in Monte Carlo (MC) simulations, it is assumed that
first the primarily produced partons from the hard scatter evolve via softer 
gluon and quark emission according to pQCD into jets of partons.  This process
continues until a cut-off $k_T$ scale ($Q_0 \sim 1$~GeV) is reached as pQCD
calculations are valid for $Q_0 >> \Lambda_{QCD}$.  After 
this phase, non-perturbative processes take over which ``cluster" the final 
partons into color singlet hadronic states via a mechanism described by
phenomenological fragmentation
models, like the Lund ``string"\cite{string} or the ``cluster"\cite{cluster}
fragmentation model.
These models usually involve quite a number of {\it a priori} unknown 
parameters that need to be tuned to the data. 

A different and purely analytical approach giving quantitative predictions of
hadronic spectra is based on the concept of ``Local Parton Hadron Duality" 
(LPHD)\cite{lphd}.  The key assumption of this hypothesis is that the 
particle yield
is described by a parton cascade where the conversion of
partons into hadrons occurs at a low virtuality scale, of the order of hadronic
masses ($Q_0 \sim 200$~MeV), independent of the scale of the primary hard 
process, and involves only low-momentum transfers; it is assumed that the 
results obtained for partons apply to hadrons as well.  
This correspondence of the partonic properties to the hadronic ones should only
be considered in an inclusive and average sense.

LPHD may be connected
to pre-confinement properties of QCD which ensure that color charges are
compensated locally\cite{amati}. 
According to the preconfinement idea,
color singlet clusters are  formed during that phase. These clusters
evolve into hadronic clusters via a smooth transformation such that
local properties remain conserved in an average sense.
With LPHD, only two essential parameters are involved in the
perturbative description: 
the effective QCD scale $\Lambda$ and a (transverse momentum) 
cut-off parameter $Q_0$, resulting in a highly constrained theoretical
framework;  non-perturbative effects are essentially reduced to 
normalization constants.
Within the LPHD approach, pQCD calculations have been carried out in the
simplest case (high energy limit) in the Double Log Approximation 
(DLA)\cite{dla1,dla2},  or in the Modified Leading Log Approximation 
(MLLA)\cite{lphd,mlla1,mlla2}, which includes higher
order terms of relative order $\sqrt{\alpha_s}$ (e.g., finite energy 
corrections) that are essential for quantitative agreement with data at present
energies.  

\section{Color Coherence}
\indent

Coherence phenomenon is an intrinsic property of QCD (and in fact of any gauge 
theory).  Its observation is important in the study of strong interactions and
in the search for deviations from the Standard Model.  

    Color coherence phenomena in the final state have been very well established
from early '80's in $e^+e^-$ annihilations\cite{cc1,cc2,cc3,cc4,cc5,cc5a}, 
in what has been
termed the ``string''\cite{cc6} or ``drag''\cite{cc7} effect.  Particle
production in the region between the quark and antiquark jets in
$e^+e^- \rightarrow q\overline{q}g$ events is suppressed.  In pQCD
such effects arise from interference between the
soft gluons radiated from the $q$, $\overline{q}$, and
$g$.  While quantum mechanical interference effects are expected in QCD, it is
important to investigate whether such effects survive the non-perturbative
hadronization process, for a variety of reactions over a broad kinematic
range, as predicted by LPHD.

It is instructive to separate the color coherence phenomena into two regions: 
the int{\underline{ra}}jet and int{\underline{er}}jet 
coherence\cite{ccgen1,ccgen2}.  
The intrajet coherence deals with the coherent effects in partonic cascades,
resulting on the average, in the angular ordering (AO) of the sequential parton
branches which give rise to the {\em hump-backed} shape of particle spectra
inside QCD jets.  The interjet coherence is responsible for the string/drag 
effect and
deals with the angular structure of soft particle flows when three or more
energetic partons are involved in the hard process.  

    The AO approximation is an important consequence of color coherence.
It results in the suppression of soft gluon radiation in the partonic cascade 
in certain regions of phase space.  For the case of outgoing
partons, AO requires that the emission angles of soft gluons decrease
monotonically as the partonic cascade evolves away from the hard process. The 
radiation is confined to a cone centered on the direction of one parton, and is
bounded by the direction of its color--connected partner.  Outside this region
the interference of different emission diagrams becomes destructive and the
azimuthally integrated amplitude vanishes to leading order.
MC simulations including coherence effects probabilistically by means 
of AO are available for both initial and final state 
evolutions.\footnote{Parton shower event generators incorporate 
AO effects in the initial
state as the time reverse process of the outgoing partonic
cascade, i.e., the emission angles increase for the incoming partons as the
process develops from the initial hadrons to the hard subprocess.}  
Another way to incorporate coherence effects in parton shower event 
generators is by using the color dipole cascade model, implemented in the
{\small ARIADNE}\cite{ariadne} MC program.

The AO is an important element of the DLA and MLLA analytic pQCD calculations,
which provides the probabilistic interpretation of soft-gluon cascades.  In fact
beyond the MLLA a probabilistic picture of the parton cascade evolution is not
feasible due to $1/N_c^2$ (where $N_c$ is the number of colors) suppressed 
soft interference contributions that appear in the 
higher-order calculations\cite{mlla1,ccgen2}.

In this report we present current experimental results on observables that test
the LPHD hypothesis in the context of color coherence phenomena from $e^+e^-$,
$ep$, and \ppbar\ collisions. 

\section{Intrajet Coherence Results}
\indent 

The study of multiparticle production in hard collision processes can yield
valuable information about the characteristic features of the partonic
branching processes in QCD and the transition from colored partons to colorless
hadrons.  
It is of great importance to know whether a smooth transition
exists between a purely perturbative regime and the soft momentum region.
Recent results on particle production are discussed in this section.

\subsection{Hump-Backed Plateau}
\indent

A striking prediction of the perturbative approach to QCD jet
physics is the depletion of soft particle production and the resulting
approximately Gaussian shape of the inclusive distribution in the variable
$\xi=\log(E_{jet}/p)=\log(1/x_p)$ for particles with momentum $p$ in a jet of 
energy $E_{jet}$---the famous ``hump-back plateau'' (see Ref. [18] and earlier
references therein).
Due to the {\em intrajet} coherence of gluon radiation (resulting on the
average in the AO of sequential branching), not 
the softest partons but those 
with intermediate energies ($E_h\propto E_{jet}^{0.3-0.4}$)
multiply most effectively in QCD cascades.

According to the expectations of the MLLA+LPHD the inclusive momentum spectra of
hadrons produced is given by:

\begin{equation}
\frac{1}{\sigma}\frac{d\sigma}{d\xi}=Const \cdot f_{MLLA}(\xi,Y,\lambda)
\label{eq1}
\end{equation}

\noindent where

\begin{equation}
Y=\log\frac{E_{jet}}{Q_0};  \lambda=\log\frac{Q_0}{\Lambda}
\label{eq2}
\end{equation}

\noindent

The function $f_{MLLA}$ is the MLLA formula for the number of final state
partons per unit $\xi$ per event and is approximately Gaussian in $\xi$.
To check the validity of the MLLA+LPHD approach, it is interesting to study the
energy evolution of the maximum, $\xi^*$, of the $\xi$ distribution.   The
prediction of the dependency of $\xi^*$ on the center of mass energy can be
expressed as (see Ref. [17] and earlier references therein):

\begin{equation}
\xi^*=Y(1/2+\sqrt{C/Y}-C/Y)+F_h(\lambda)
\label{eq3}
\end{equation}

\noindent where

\begin{equation}
C=\left(\frac{11N_c/3+2n_f/3N_c^2)}{4N_c}\right)^2\cdot\frac{N_c}{11N_c/3-2n_f/3}
\label{eq4}
\end{equation}

\noindent 
with $n_f$ the number of quark flavors and
$F_h(\lambda)$ a function that depends on the hadron type, $h$, through the 
ratio $\lambda$:

\begin{equation}
F_h(\lambda)=-1.46\lambda+0.207\lambda^2 \pm 0.06
\label{eq5}
\end{equation}

The shapes of  the measured particle energy spectra in $e^+e^-$ annihilation
from early '90's have been surprisingly close, over the whole 
momentum range  down to momenta of a few hundred MeV, to the MLLA+LPHD
predictions.
These observations can be taken as evidence that the perturbative phase
of the cascade development indeed leaves its imprint on the
final state hadrons supporting the hypothesis that color coherence effects
survive the hadronization process as suggested by LPHD.

Recently,  new  data on  charged  particle   spectra,  exploring  higher energy
regimes,  have  become   available  from LEP,  HERA,  and   TEVATRON.  The HERA
experiments    concentrate  on the   ``current"   fragmentation   region in DIS
(fragmentation products of the  outgoing quark) and perform the analysis in the
Breit frame, where the exchanged boson is completely spacelike. The DIS current
fragmentation   functions  at a  momentum  transfer  $Q$ are   analogous to the
$e^+e^-$ fragmentation functions at center of mass energy equal to 
$Q$\cite{unive}. The new
data confirm with much increased statistical significance the features observed
in   $e^+e^-$:    approximately  Gaussian   shape of  the  $\xi$   spectra with
peak-position and width increasing with $Q$
as predicted in MLLA.  

\begin{figure}[htb]
  \centering
\mbox{
\subfigure[Evolution of the $1/Ndn/d\log(1/x_p) $ 
           distributions with $Q$.  The curves are MLLA fits. ]
{\psfig{figure=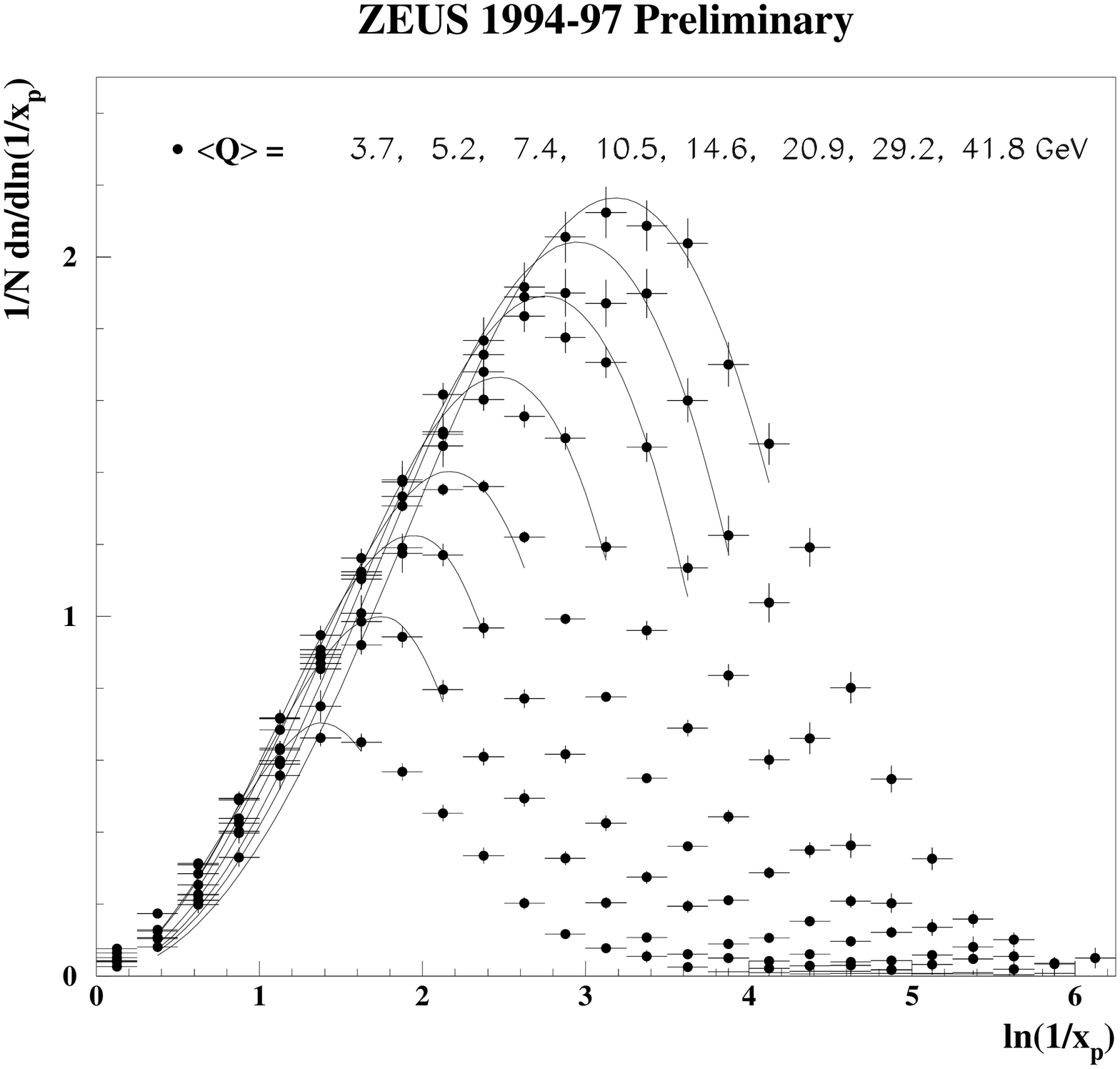,width=.45\textwidth}}\quad
        \subfigure[Evolution of the peak position $log(1/x_p)_{max}$ with $Q$.]
{\psfig{figure=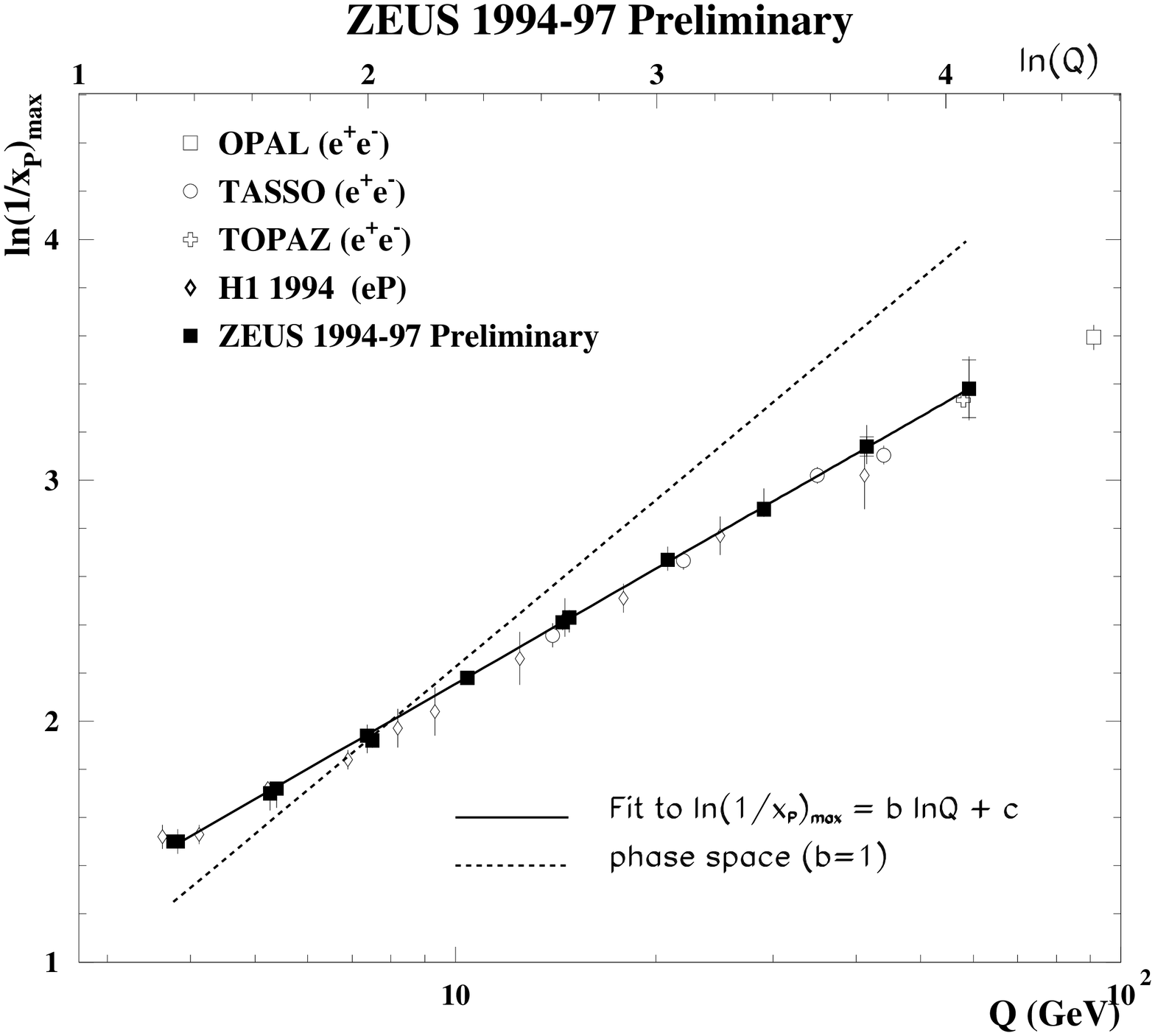,width=.45\textwidth}} 
}
  \caption[]{Comparison of preliminary ZEUS inclusive charged particle
             momentum distributions
             with MLLA predictions and $e^+e^-$ data.}
  \label{zeus_frag}
\end{figure}

Figure~\ref{zeus_frag}a shows preliminary $\xi$ distributions for charged 
particles in
the current fragmentation region of the Breit frame as a function of $Q$ from
ZEUS.  These distributions are approximately Gaussian in shape with mean
charged multiplicity given by the integral of the distributions.  As $Q$
increases the multiplicity increases and the peak of the distribution shifts to
larger values of $\xi$ (i.e., smaller values of momenta).
Figure~\ref{zeus_frag}b shows this peak position, $\log(1/x_p)_{max}$, as a
function of Q for the ZEUS data and of $\sqrt{s}$ for the $e^+e^-$ data.  Over
the range shown the peak moves from $\approx$~1.5 to 3.3.  The ZEUS data points
are consistent with those from TASSO and TOPAZ and a clear agreement in the
rate of growth of the ZEUS points with the OPAL data at higher $Q$ is observed.
From Fig.~\ref{zeus_frag}b it is clear that the ZEUS data are incompatible
with the assumption of an incoherent branching picture or a
simple phase-space model ($\xi^* \approx Y + const$).  ZEUS also fitted the MLLA
predictions (eq.~\ref{eq3}) to their $\xi^*$ evolution data, assuming 
$Q_0=\Lambda \equiv Q_{eff} \equiv \Lambda_{eff}$, and extracted a value of 
$\Lambda_{eff} \approx 245$~MeV
which is in agreement with a similar value obtained from H1, CDF, or combined 
$e^+e^-$ data.  Notice that decreasing $Q_0$ means extending the responsibility
of the perturbative stage beyond its formal range of applicability.  However,
in MLLA calculations this limit is smooth yielding a finite result even if the
coupling gets arbitrarily large.

\begin{figure}[htb]
 \vspace{-0.2cm}
 \centerline{\psfig{figure=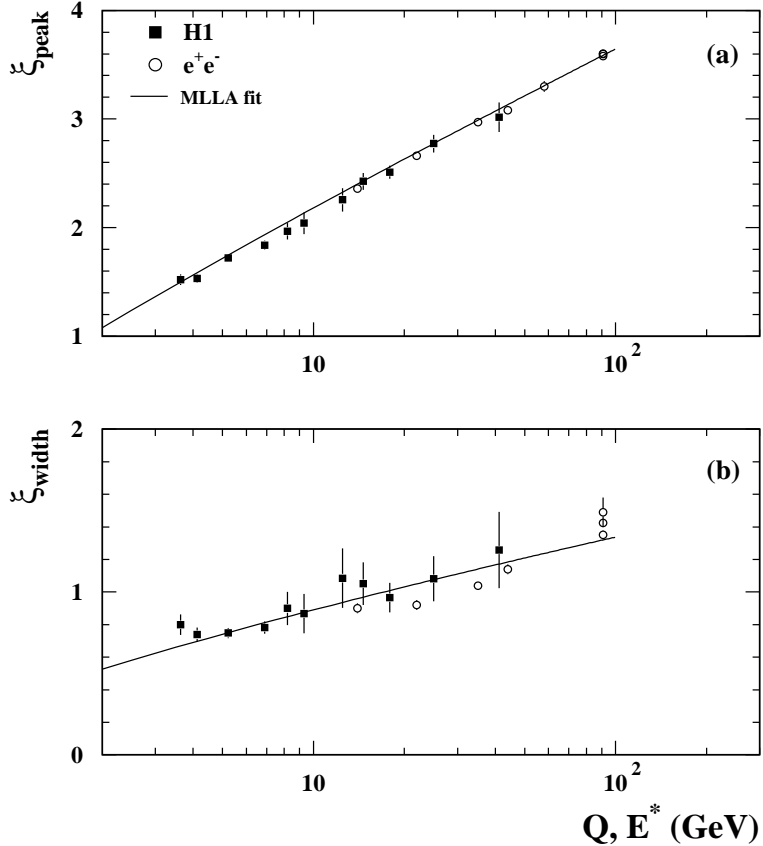,width=10cm}}
 \vspace{0.6cm}
 \caption[]{H1 results (solid symbols) showing the evolution of (a) the peak
            and (b) the width of the fragmentation function as a function of
            $Q$ compared to $e^+e^-$ results (open symbols) as a function of
            the center of mass energy, $E^*$.  The solid line is a fit to MLLA
            expectations.}
 \label{h1_frag}
 \vspace{0.1cm}
\end{figure}

\begin{figure}[htb]
  \centering
\mbox{
\subfigure[$1/Ndn/dlog(1/x_p)$ distribution at 183 GeV.
           The curves are MLLA fits with different Gaussian parametrizations. ]
{\psfig{figure=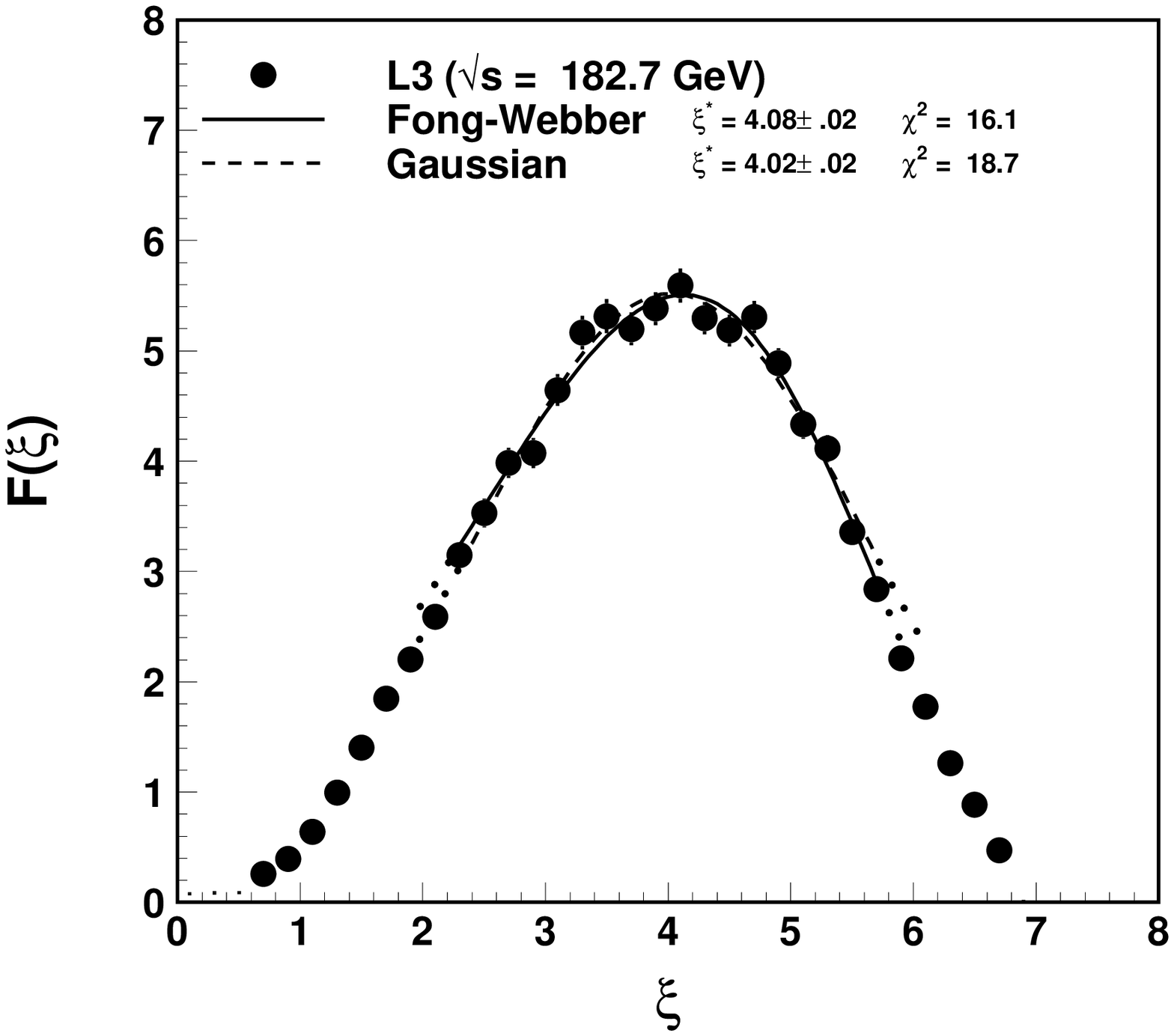,width=.45\textwidth}}\quad
        \subfigure[Evolution of the peak position $\xi^*$ 
        with energy.  The dashed line is a MLLA fit and the solid line is a fit
        to DLA.  Clearly DLA disagrees with the data. ]
{\psfig{figure=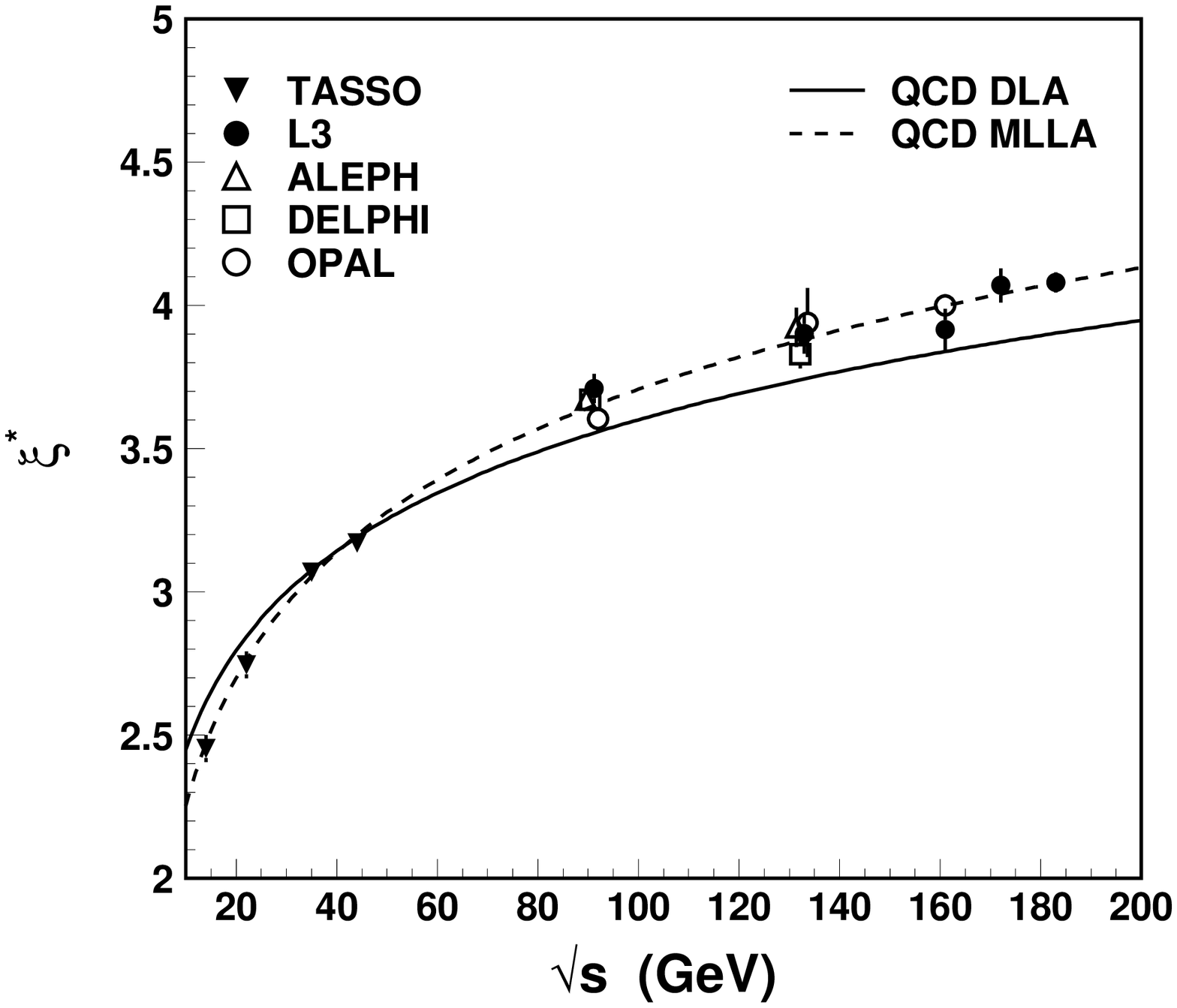,width=.45\textwidth}} 
}
  \caption[]{Comparison of preliminary L3 inclusive charged particle 
            momentum distributions
             with MLLA predictions}
  \label{l3_frag}
\end{figure}

Figures~\ref{h1_frag} and \ref{l3_frag} show similar distributions from the H1
and L3 experiments.  A simultaneous MLLA fit to the peak and width values 
obtained from the present H1 data alone yields a value of 
$\Lambda_{eff} = 0.21 \pm 0.02$~GeV\cite{h1_frag}.  
From Fig.~\ref{l3_frag}b we see that 
the DLA calculations clearly disagree with the $e^+e^-$ data.  The finite
energy corrections included in the MLLA predictions seem important.  We also
notice that the value of $\xi^*$ at the $Z^0$ corresponds to the rather low
momenta $x_p \approx 0.02$ or $p\approx 1$~GeV (Fig.~\ref{l3_frag}a).

\begin{figure}[htb]
  \centering
\mbox{
\subfigure[Evolution of $\xi$ with jet opening angle, $\Theta$,
           for $M_{\rm JJ}$~=~390~GeV.]
{\psfig{figure=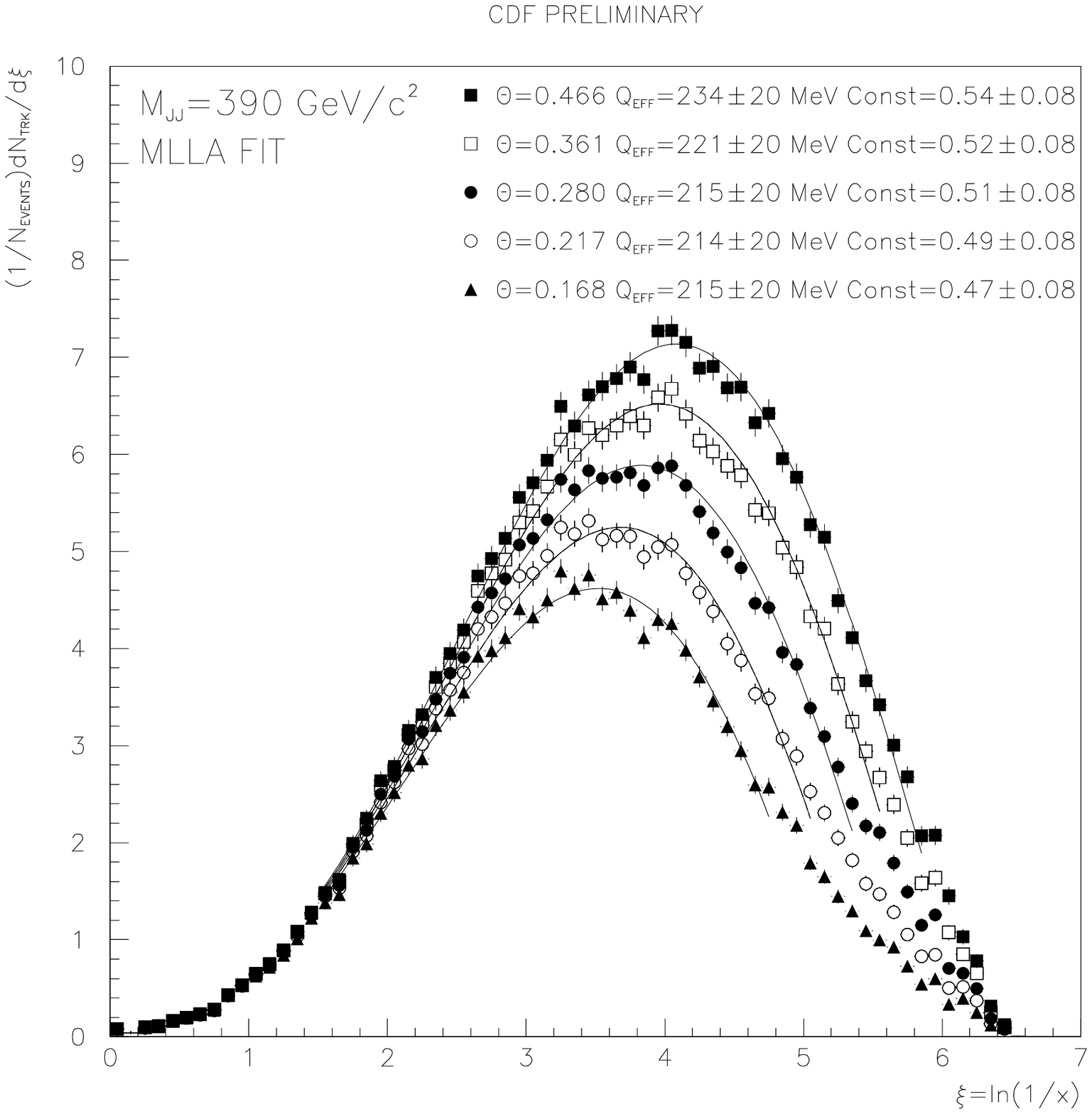,width=.45\textwidth}}\quad
        \subfigure[Evolution of the peak position with $M_{\rm JJ}\Theta$.]
{\psfig{figure=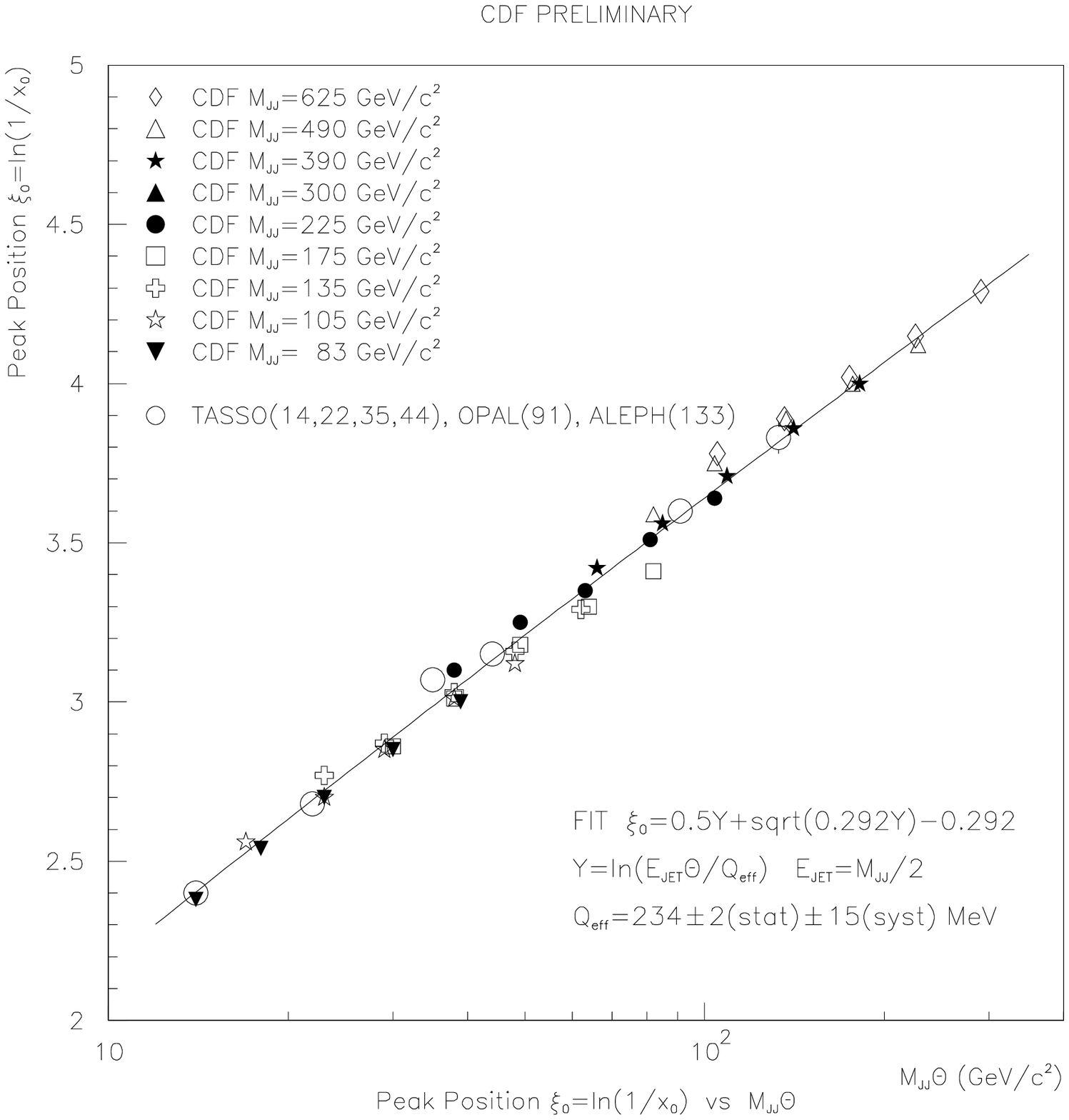,width=.45\textwidth}} 
}
  \caption[]{Comparison of preliminary CDF inclusive momentum distributions
             with MLLA predictions and $e^+e^-$ annihilation data.}
  \label{cdf_frag}
\end{figure}

Confirmation of the  MLLA+LPHD approach has also been presented 
by CDF. This experiment studies charged particle momentum
distributions in subsamples of dijet events. For fixed  dijet masses
in the range $83<M_{\rm JJ}< 625$ GeV/c$^2$,
the $\xi$ distribution of tracks, within cones of various opening 
angle $\Theta$ (with respect to the jet axis), is 
studied (see Fig.~\ref{cdf_frag}a).
As dijet mass $\times$ jet opening angle
increases, the peak of the spectrum, $\xi_o$,
shifts towards larger values
of $\xi$  in perfect agreement with the 
$e^+e^-$ data, as shown in Fig.~\ref{cdf_frag}b.
The MLLA fit (superimposed in Fig.~\ref{cdf_frag}b) is in excellent agreement
with the data and yields $Q_{eff}=234 \pm 2(stat) \pm 15(syst)$~MeV, confirming
that in this approximation the domain of pQCD extends down to 
$Q_{eff} \sim \Lambda_{QCD}$.
Similar analyses should be possible in DIS and photoproduction
at HERA but have not yet been attempted.

\subsection{Identified Particle Spectra}
\indent

The $\xi$-spectra for a variety of identified particles/resonances has been
studied at the $Z^0$ pole at SLAC and at LEP.  Recently the SLD Collaboration
has reported measurements of the differential production cross sections as a
function of $x_p$ of several identified hadron and antihadron species in
inclusive hadronic $Z^0$ decays, as well as separately for $Z^0$ decays into
light ($u$, $d$, $s$), $c$, and $b$ flavors\cite{sld}.  These results have been 
compared to MLLA as well as the predictions of three fragmentation models.

\begin{figure}[htb]
 \vspace{-0.2cm}
 \centerline{\psfig{figure=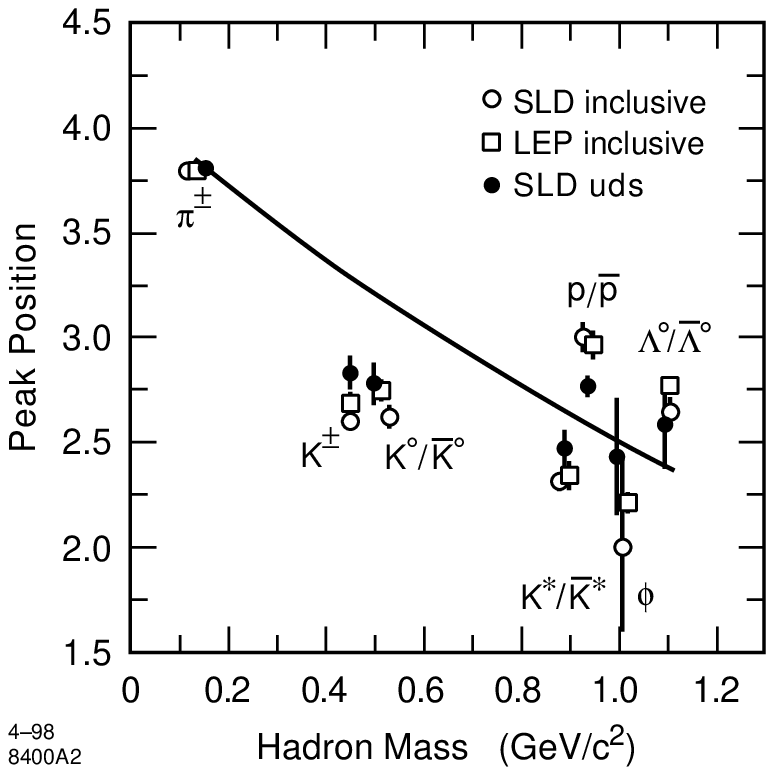,width=8cm}}
% \vspace{0.9cm}
 \caption[]{Peak positions $\xi^*$ from fits to the $\xi$ distributions in
 flavor-inclusive and light-flavor hadronic $Z^0$ decays.  Also shown are the
 averages of similar flavor-inclusive results from experiments at LEP.  The
 line is the result of an ad hoc exponential fit to the SLD light-flavor data.}
 \label{sld_frag}
 \vspace{0.1cm}
\end{figure}

The fitted peak positions $\xi^*$ for the various particles are plotted as a
function of the hadron mass in Fig.~\ref{sld_frag}, along with averages of
similar measurements from experiments at LEP\cite{bohrer}, with which they are
consistent.  The distribution for pions peaks at a higher $\xi$ value than
those of the other hadron species, but otherwise there is no obvious 
mass-dependence.\footnote{One may also argue that Fig.~\ref{sld_frag}
indicates that 
the peak position $\xi^*$ decreases as a function of mass
differently for baryons and mesons.  These results may provide additional
insights concerning the LPHD concept.}
We notice that within the framework of the MLLA+LPHD picture there is no recipe
for relating the $Q_0$ cut-off parameter to the masses of the produced hadrons
and their quantum numbers.  

The DELPHI Collaboration has recently reported preliminary results on the
production of charged and neutral kaons, protons, and $\Lambda$s at center of 
mass energies above the $Z^0$ pole.\cite{angelis}.  The data are found to be 
in good agreement
with the MLLA predictions.  This comparison confirms the perturbative
expectation (eq.~\ref{eq3}) that for different particle species the energy 
dependence of $\xi^*$ is universal.  
The fit of the data points to expression (\ref{eq3}), where $F_h(Q_0)$ was 
taken as a free
parameter and $\Lambda$ was fixed to 150 MeV (this value comes from the
description of the pion spectra with $\Lambda=Q_0$) yielded a value for 
$Q_0$ of about 330~MeV consistent for the different particle types. 

\subsection{Invariant Energy Spectrum}
\indent

The analytical perturbative approach allows one to predict the limit of
the  one-particle invariant  density in QCD jets  $E dn/{d^3p}\equiv
dn/dy\,d^2{\bf p_T}$ at very small momenta $p$ or, equivalently, in the
limit of vanishing rapidity and transverse momentum\cite{ccgen1}. 
If the dual description of hadronic and partonic states is 
adequate down to very small momenta, a finite, energy-independent limit of
the invariant hadronic density, $I_0$, is expected. 
This is a direct consequence of the color coherence
in soft gluon branching. 
Indeed, long wavelength gluons are emitted by the total conserved color 
current, which is independent of the internal structure of a jet and its energy.
A possible rise  of $I_0$  with center-of-mass energy
would indicate that either coherence or the LPHD
(or both) break down. Since color coherence is a general property of QCD as a
gauge theory, it is the LPHD concept that is tested in  measurements of the
soft hadrons.

The $e^+e^-$ annihilation data on charged and identified
particle inclusive spectra have been found to 
follow the MLLA  prediction surprisingly well, also at
low center-of-mass energies. The invariant spectra at low momentum
scale approximately (within 10\%) between 1.6 and 161~GeV
and agree with  perturbative calculations  which become very sensitive
to the strong running of $\alpha_S$ at small scales~\cite{lo}. 

The H1 Collaboration has reported the first Breit frame measurements
of the invariant energy spectra in DIS as a function of $Q$\cite{h1_frag}. 
For sufficiently high $Q$, the data  show that the low-momentum
limit in that region of phase space
is essentially independent of $Q$ and indeed similar 
to that in $e^+e^-$ annihilation.  

\section{Interjet Coherence Results in \ppbar\ Interactions}
\indent

    The study of coherence effects in hadron--hadron collisions is considerably
more subtle than that in $e^+e^-$ annihilations due to the presence of colored
constituents in both the initial and final states.  
During a hard interaction, color is transferred from one
parton to another and the color--connected partons act as color antennae, with
interference effects taking place in the initial or final states, or between 
the initial and final states.
Gluon radiation from the incoming and outgoing partons forms jets of hadrons
around the direction of these
colored emitters.  The soft gluon radiation pattern accompanying any hard
partonic system can be represented, to leading order in $1/N_c$ 
as a sum of contributions corresponding to the color--connected
partons.  Within the perturbative calculations, this is a
direct consequence of interferences between the radiation of various
color emitters, resulting in the QCD coherence
effects \cite{cc7,ccgen2,cc8}.

\subsection{Multijets}
\indent

Both the CDF\cite{cc10} and D\O\cite{d0_coh} Collaborations have measured
spatial correlations between the softer third jet and the second leading-$E_T$
jet in $p\overline{p}~\rightarrow~3jets~+~X$ events
to explore the initial-to-final state coherence effects in \ppbar\
interactions.

In the \D0\ analysis the jets were reconstructed using a fixed-cone clustering
algorithm with 0.5 cone radius.  The corrected
transverse energy of the highest-\et\ jet of the event was required to be
above 115 GeV while the third jet was required to have $E_T > 15$ GeV.
The interference between the second and the third jet is displayed using the
polar variables $R=\sqrt{(\Delta\eta)^2 + (\Delta\phi)^2}$ and
$\beta = \tan^{-1}(\frac{sign(\eta_2)\cdot\Delta\phi}{\Delta\eta})$; where
$\Delta\eta = \eta_3 - \eta_2$
and $\Delta\phi = \phi_3 - \phi_2$, in a search
disk of $0.6 < R < \frac{\pi}{2}$. (Pseudorapidity $\eta=-\ln[\tan(\theta/2)]$, 
where $\theta$ is the polar angle of the jet with respect to the proton beam.)

\begin{figure}[tb]
 \vspace{-0.2cm}
 \centerline{\psfig{figure=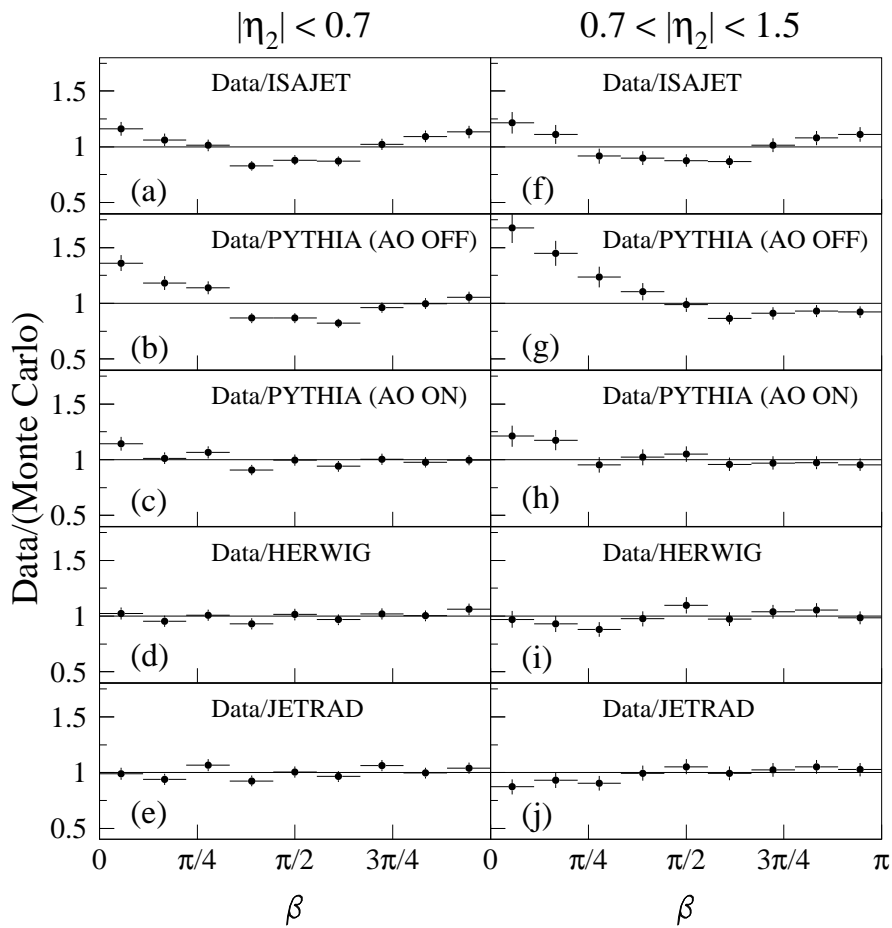,width=13cm}}
 \vspace{0.6cm}
% \vspace{2cm}
 \caption[]{Ratio of $\beta$ distributions between data and the predictions of:
            (a) {\small {ISAJET}}, (b) {\small {PYTHIA}} with AO off,
            (c) {\small {PYTHIA}} with AO on, (d) {\small {HERWIG}},
            (e) {\small {JETRAD}} for the central region; and (f)-(j) for the
            forward region respectively. The error bars
            include statistical and uncorrelated systematic uncertainties.}
 \label{PL_FIG4}
 \vspace{0.1cm}
\end{figure}

Figure~\ref{PL_FIG4} shows the ratio of the $\beta$ distributions
for the \D0\ data relative to several MC predictions for both
central ($|\eta_2| < 0.7$) and forward ($0.7 < |\eta_2| < 1.5$) regions.
Detector position and energy resolution effects have been included in the
MC simulations. 
The absence of color interference effects in
{\small ISAJET}\cite{isajet} results in a disagreement with the \D0\ $\beta$
distributions.  The
data show a clear enhancement of events compared to {\small {ISAJET}} near the
event plane (i.e., the plane defined by the directions of the second jet and
the beam axis, $\beta=0,~\pi$) and a depletion in the transverse plane
($\beta=\frac{\pi}{2}$).
This is consistent with the expectation from initial--to--final state color
interference that the rate of soft jet emission around the event plane be
enhanced with respect to the transverse plane.  
However, {\small {HERWIG}} 5.8\cite{cluster} which contains initial and final 
state interference effects implemented by means of the AO approximation of
the parton cascade, agrees well with the data.  The \D0\ data have also been
compared to {\small {PYTHIA}} 5.7\cite{pythia} which also simulates the color
interference effects
with the AO approximation.  The {\small {PYTHIA}}
predictions include string fragmentation.  Without AO the {\small {PYTHIA}}
distributions are significantly different from the data, while with AO turned
on there is much better agreement, although there are still some residual
differences in the ``near beam'' region.
Finally, the ${\cal O}(a_{s}^{3})$ tree-level QCD prediction from 
{\small JETRAD}\cite{jetrad}
describes the coherence effects seen in the data in both $\eta$ regions.

\newpage
\subsection{$W$+Jets}
\indent

    In  $p\overline{p} \rightarrow W+Jets$ events, the angular distribution of 
soft gluons about the 
colorless $W$ boson is expected to be uniform, while the distribution around 
the jet is expected to have structure due to the colored partons in the jet.
\D0\ studies these effects by comparing the
distributions of soft particles around the $W$ boson and opposing
jet directions. This
comparison reduces the sensitivity to global detector and underlying
event biases that may be present in the vicinity of the $W$ boson
and the jet.

Once the $W$ boson direction has been determined in the \D0\ detector, the
opposing jet is identified by selecting the leading--$E_T$ jet in the
$\phi$ hemisphere opposite to the $W$ boson.  Annular regions are drawn
around both the $W$ boson and the jet in $(\eta,\phi)$ space.
The angular distributions of towers 
($\Delta\eta\times\Delta\phi = 0.1\times0.1$) above the 250~MeV threshold are
measured in these annular regions using the polar variables
$R=\sqrt{(\Delta\eta)^2 + (\Delta\phi)^2}$ and $\beta_{W,Jet} =
\tan^{-1}(\frac{sign(\eta_{W,Jet})\cdot\Delta\phi_{W,Jet}}
{\Delta\eta_{W,Jet}})$; where $\Delta\eta_{W,Jet} = \eta_{Tower} -
\eta_{W,Jet}$ and $\Delta\phi_{W,Jet} = \phi_{Tower} - \phi_{W,Jet}$, in a
search disk of $0.7 < R < 1.5$.  Color coherence effects are expected
to manifest themselves as an enhancement in the energetic tower
distribution around the tagged jet in the event plane relative to
the transverse plane (when compared with the $W$ boson distribution).  
These coherence effects are similar to the string/drag effects observed in the 
$e^+e^- \rightarrow q\overline{q}g$ events.  

\begin{figure}[tb]
  \centering
\mbox{
\subfigure[Jet/$W$ tower multiplicity ratio as a function of $\beta$.]  
{\psfig{figure=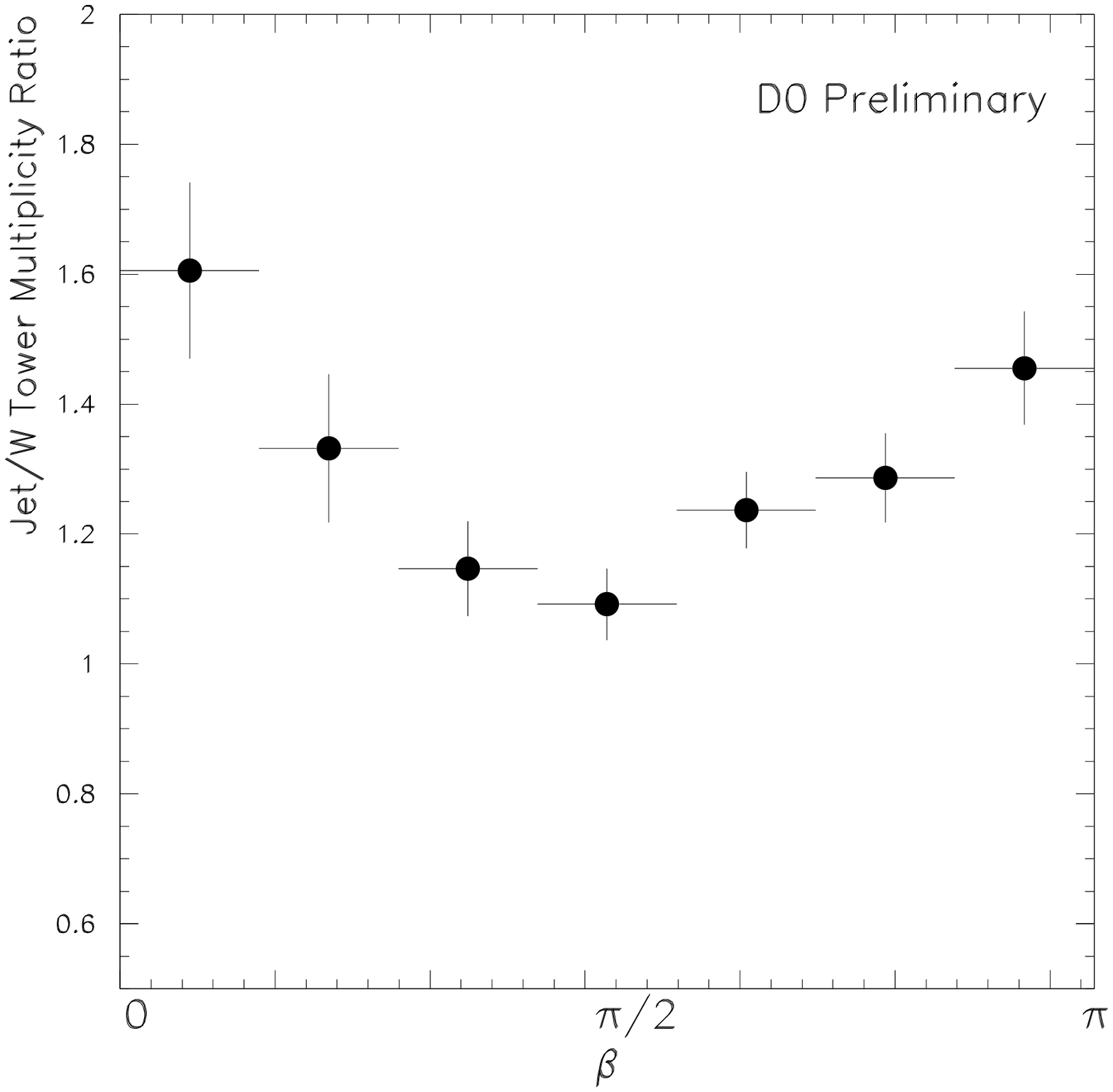,width=.45\textwidth}}\quad
        \subfigure[Ratio of event plane to transverse plane of
          Jet/$W$ tower multiplicity for \D0\ data, {\small PYTHIA} with
          various coherence implementations, and a MLLA QCD calculation.
          The errors are statistical only.]
{\psfig{figure=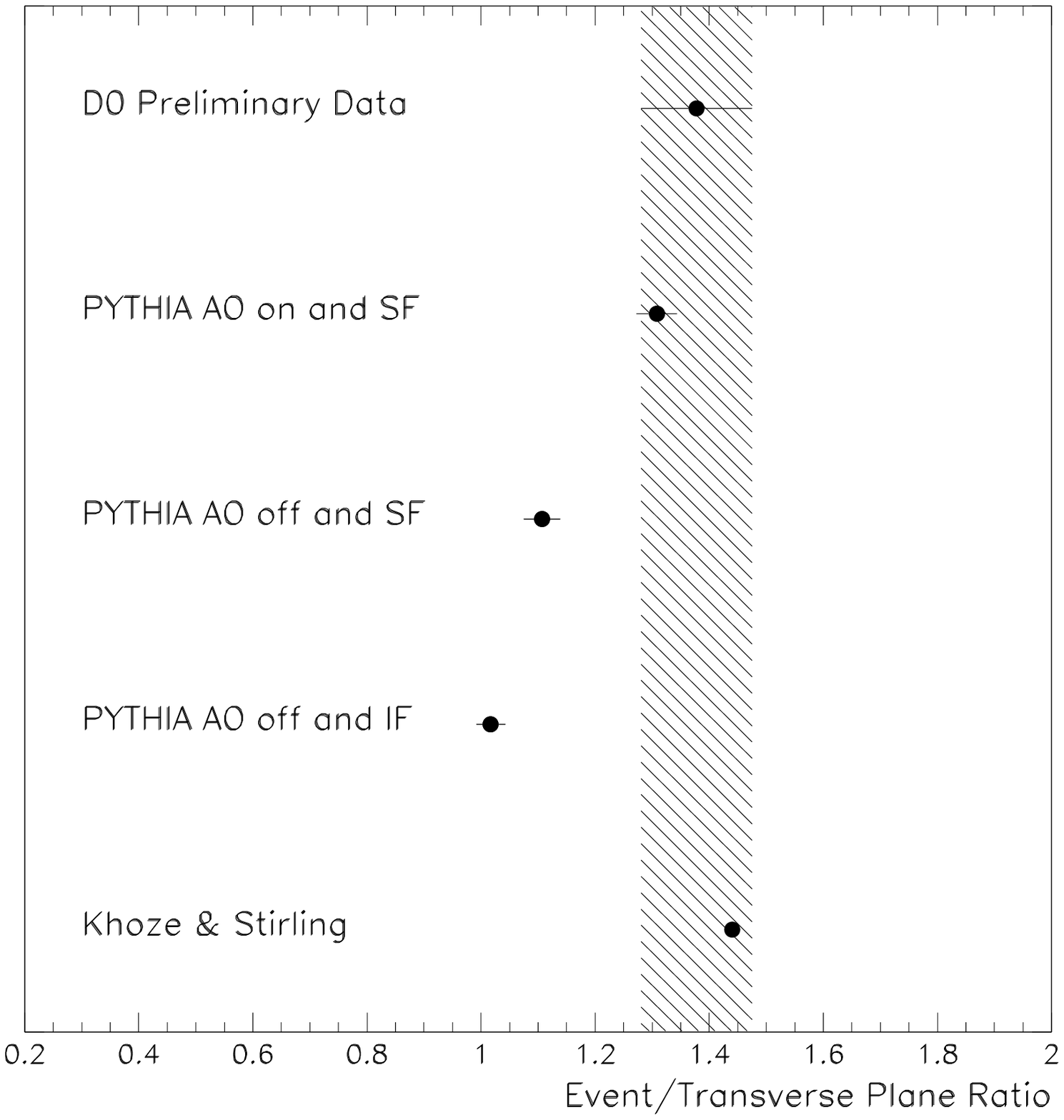,width=.45\textwidth}} 
}
  \caption[]{\D0\ preliminary results on \wjets\ coherence.}
  \label{cc_wjet}
\end{figure}

\D0\ has analyzed $W \rightarrow e+\nu$ events requiring at least one jet with 
$E_T > 10$~GeV,  $|\eta_{Jet}|<0.7$, and 
$W$ boson rapidity $|y_W|<0.7$.
Additionally, the $z$ component of the
event vertex was restricted to $|z_{vtx}|<20$~cm to retain the
projective nature of the calorimeter towers.

The data angular distributions are compared to three MC samples, 
generated with different levels of color coherence effects, using the 
{\small PYTHIA} 5.7 parton shower event generator and passed 
through a full detector simulation.
{\small PYTHIA}, with
both AO and string fragmentation (SF) implemented, accounts for color
coherence effects at both the perturbative and non-perturbative
levels.  Turning off AO removes the perturbative contribution, and
using independent fragmentation (IF) eliminates the non-perturbative
component.  Finally, a comparison to a MLLA+LPHD pQCD calculation of Khoze and 
Stirling\cite{cc_khoze_stirling}
is also presented.

Figure~\ref{cc_wjet}a shows the ratio of the tower multiplicity
around the jet to the tower multiplicity around the $W$
as a function of $\beta$.  The number of
towers is greater for the jet than for the $W$ boson and the excess is
enhanced in the event plane ($\beta = 0, \pi$) and minimized in the
transverse plane ($\beta = \frac{\pi}{2}$), consistent with the
expectation from initial--to--final state color interference effects.  The
errors include only statistical uncertainties, which are significantly
larger than all systematic uncertainties considered.  

A measure of the observed color coherence effect is obtained by
calculating the Jet/$W$ tower multiplicity enhancement of the event
plane ($\beta = 0,\pi$) to the transverse plane ($\beta = \pi/2$), which would 
be expected to be unity in the absence of color coherence effects.
This ratio of ratios is insensitive to the overall normalization of
the individual distributions, and MC studies have shown that
it is relatively insensitive to detector effects.
Figure~\ref{cc_wjet}b compares the data to the various
{\small PYTHIA} predictions and to the MLLA+LPHD calculation.  There is good 
agreement with
{\small PYTHIA} with AO on and string fragmentation, and disagreement
with AO off and string fragmentation or AO off and independent
fragmentation.  These comparisons imply 
that for the process under study, string fragmentation alone cannot 
accommodate the effects seen in the data.  The AO approximation is an element 
of parton-shower event generators that needs to be included if color 
coherence effects are to be modeled successfully.
Finally, the analytic predictions by Khoze and Stirling are in 
agreement with the data, giving additional evidence supporting the 
validity of the LPHD hypothesis.

\section{A Color ``Reconnection" Study in Hadronic $Z^0$ Decays}
\indent 

Most implementations of QCD coherence effects are based on a probabilistic
scheme (e.g., AO approximation) where interference terms of relative order 
$1/N_c^2$ are ignored (the so-called large $N_c$ approximation).  In this
picture the way the partons are connected to form color singlet states is
uniquely specified.  For example, in $Z^0 \rightarrow q\bar{q}gg$ events, in
which two gluons are radiated against a $q\bar{q}$ pair from a $Z^0$ decay, the
quark is color-connected to one of the gluons, this gluon is connected to the
second gluon, and the second gluon is connected to the antiquark.  Thus, the
entire event consists of a color singlet state.

Beyond the large $N_c$ approximation, the color configurations that the partons
follow when connected to each other is not longer specified uniquely.  In our
example, the possibility that the $q$ and $\bar{q}$ form a color singlet by
themselves, with the two gluons $gg$ forming a separate color singlet, is
suppressed by a factor of $1/(N_c^2-1)$ relative to the leading configuration
described above.
The possibility to connect the partons in this latter manner is often called
color ``reconnection" or recoupling.  Color ``reconnection" is, however, an 
unfortunate terminology since the partons have not been 
physically {\bf re}-connected.  

Color reconnection effects are of fundamental importance for our understanding
of the confinement mechanism.  Does Nature select a particular configuration at
random or some configuration is dynamically favored in forming color singlet
states?  Recently there has been a lot of interest on color reconnection due to
the possibility that higher order color rearrangement diagrams could affect the
$W$ mass measurement at LEP-II.  So far there has not been any experimental
evidence on this expectation suggesting that these effects might be 
small in $WW$ pair events\cite{monica}.

\begin{figure}[tb]
 \vspace{-0.2cm}
 \centerline{\psfig{figure=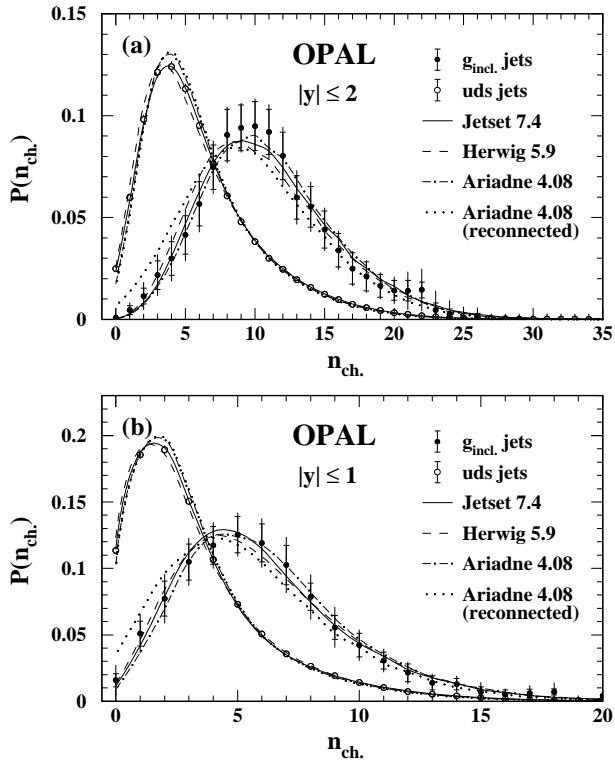,width=8cm}}
 \vspace{0.6cm}
 \caption[]{OPAL preliminary results on charged particle multiplicity in the
            rapidity intervals (a) $|y| \leq 2$ and (b) $|y| \leq 1$ for 
            41.6~GeV ``$g_{inc}$ jets" and 45.6~GeV uds quark jets.  The total
            uncertainties are shown by vertical lines.  The statistical
            uncertainties are indicated by horizontal bars.  The predictions of
            various parton shower MC event generators are also shown.}
 \label{opal_reco}
 \vspace{0.1cm}
\end{figure}

The OPAL Collaboration has performed a search for color reconnection effects in
$Z^0 \rightarrow q\bar{q}g_{inc}$ events, in which the $q$ and $\bar{q}$ are
identified quark (and antiquark) jets which appear in the same hemisphere of
an event.  The object $g_{inc}$, taken to be the $g$ jet, is defined by all
particles observed in the hemisphere opposite to that containing the $q$ and
$\bar{q}$ jets.  In the limit that the $q$ and $\bar{q}$ are collinear the
gluon jets are produced from a color singlet point source, corresponding to the 
definition of gluon jets in the theoretical calculations.  

The {\small ARIADNE} MC program with and without reconnection effects is 
compared to the OPAL preliminary data.  The version of {\small ARIADNE} 
with
reconnection predicts fewer (more) particles at small (large) rapidities and 
energies than are observed in either the data or other standard MC programs.
Furthermore, {\small ARIADNE} with reconnection
predicts a smaller charged particle multiplicity for the $g_{inc}$
hemisphere than the other standard MC simulations or the multiplicity measured 
in data (see Fig.~\ref{opal_reco}).  Figure~\ref{opal_reco}
also shows the multiplicity distributions for 
light quark ($u,d,s$) jets in $Z^0 \rightarrow q\bar{q}$ decays, which appear
to be insensitive to reconnection effects.

OPAL performed two quantitative tests to assess the difference between the data
and the predictions of {\small ARIADNE}.  The first test is based on
the comparison of the ratio of the mean gluon to light quark jet charged 
particle multiplicity between the data and the various MC predictions, 
and the second one is based on the probability for a $g_{inc}$ jet to have 
five or fewer
charged particles with $|y| \leq 2$.  Both tests showed that the 
{\small ARIADNE} reconnection model is disfavored by the data, whereas the
standard QCD MC programs ({\small HERWIG}, {\small JETSET}, and 
{\small ARIADNE} without reconnection) reproduce the experimental results well.

\section{Conclusions}
\indent 

The high precision data from HERA, LEP, SLD, and TEVATRON
have provided a detailed testing ground for the strong interactions.  Beautiful
agreement of the data, from  $e^+e^-$, $ep$, and \ppbar\ collisions,
with analytic pQCD calculations, based on MLLA and LPHD,
has been seen in several inclusive observables sensitive to color coherence 
phenomena.  Although the support of the LPHD from the current data is strong, 
it is important to continue investigating the limitations of this picture,
since it is not clear {\it a priori} for which observables and in which 
kinematic regions it applies.  Finally, the traditional parton shower event 
generators which incorporate color coherence effects at the perturbative and 
non-perturbative stages seem to describe well the interjet coherence phenomena
observed in \ppbar\ collisions.

\section{Acknowledgements}
\indent 

I express my deep appreciation to Valery Khoze for numerous valuable
discussions on this work.  I would also like to thank 
Alessandro De Angelis, Phil Burrows, Tony Doyle, Martin Erdmann, Costas Foudas, 
Bill Gary,  Rick Van Kooten, Andrey Korytov, Joachim Mnich, and David Muller who
provided me with material for this review.  Special thanks to Valery Khoze,
Hugh Montgomery, and Gregory Snow for their comments on this manuscript.
\end{document}